\documentstyle[seceq,epsf,wrapfig,twoside]{ptptex}
\notypesetlogo  %comment in if to eliminate PTPTeX logo
\title{Hadronic Weak Decays of ${\mib \Lambda}_b$ Baryon\\ in the 
Covariant Oscillator Quark Model }
\author{%
Rukmani {\sc Mohanta}, Anjan K. {\sc Giri}, Mohinder P. {\sc Khanna},\\
Muneyuki {\sc Ishida},$^{*}$ Shin {\sc Ishida}$^{**}$
and Masuho {\sc Oda}$^{***}$
}
\inst{%
Department of Physics,
Panjab University, Chandigarh - 160014, India\\
$^{*}$Department of Physics, Tokyo Institute of Technology\\
Tokyo 152-8551, Japan\\
$^{**}$Atomic Energy Research Institute, 
College of Science and Technology\\
Nihon University, Tokyo 101-0062, Japan\\ 
$^{***}$Faculty of Engineering, Kokushikan University, 
Tokyo 154-8515, Japan\\
}
\recdate{%
December 13, 1998
}
\abst{%
Cabibbo allowed two-body hadronic weak decays of $\Lambda_b$
baryons are analyzed in the factorization approximation.
We use the covariant oscillator quark model to evaluate the
heavy $\to $ heavy and heavy $\to $ light form factors.
When applied in the heavy quark limit, our form factors 
satisfy all the constraints imposed by heavy quark symmetry.
The decay rates and up-down asymmetries for $\Lambda_b$
baryon decaying into $\Lambda_c +P(V)$ are calculated.
It is found that the up-down asymmetry is negative in all
these decay modes. Furthermore, the prediction
$Br(\Lambda_b \to \Lambda+J/\psi )=2.49 \times 10^{-4}$
is consistent with the recent experimental data. Finally
it is pointed out that the CKM-Wolfenstein parameter
$\rho^2+\eta^2 $, where $\eta$ is the $CP$ phase, can be
determined from the ratio of the widths $\Lambda_b \to
\Lambda \bar D^0 $ and $\Lambda_b \to \Lambda J/\psi $,
independent of the QCD parameter. The value of $(\rho^2
+\eta^2)^{1/2} $ calculated in our model agrees very
well with the value recently predicted by Rosner.
}
\begin{document}
\maketitle

\setcounter{tocdepth}{4}

\section{Introduction}

Progress in understanding nonleptonic weak
decays of bottom baryons has been very slow both theoretically and
experimentally. While some new data of charmed baryon nonleptonic
weak decays have become available, the experimental situation
for hadronic weak decays of bottom baryons is meagre. The
decay mode $\Lambda_b \to \Lambda J/\psi $ is the first
successful measurement \cite{ref00} of the exclusive
hadronic decay rate of bottom
baryons. In the near future, one can expect
new data on exclusive bottom baryon decays calling for a
comprehensive theoretical analysis of these decay modes.
These decay processes can provide useful information for QCD
effects in weak decays and indirect CP asymmetry which
involve the CKM-Wolfenstein parameters $\rho $ and $\eta $. However,
a rigorous and reliable approach suitable for analyzing these
decays does not exist at this time, and one must rely on some
approximation to deal with them. The well-known factorization
approach,\cite{ref1} which has been applied successfully
to nonleptonic $B$ meson decays \cite{ref2}
can also be applicable to bottom baryon decays.
In this approximation the effects of final state interactions (FSI)
are neglected. These interactions are thought to be less important in
bottom baryon decay, since the decay particles in the two-body 
final state are energetic and moving fast, allowing little
time for significant final state interactions.

In previous papers, we made
analyses of exclusive semi-leptonic\cite{ref6p} and 
non-leptonic\cite{Bmeson} decays of $B$ mesons
 by using the
covariant oscillator quark model(COQM),\cite{ref3}
leading to satisfactory results.
In this paper we discuss the nonleptonic decays $\Lambda_b \to
\Lambda_c P(V) $ and $\Lambda_b \to \Lambda J/\psi $ using 
this model.
One of the most important motives for 
COQM is to covariantly describe the centre-of-mass 
motion of hadrons, retaining the considerable successes of
the non-relativistic quark model on the static properties of
hadrons. As a result, in COQM we can treat both static and non-static
problems simultaneously. A common systematic treatment of the systems
with any quark flavor configuration is another feature of COQM. 
A key point in COQM for doing this is to treat directly
the squared masses of hadrons in contrast to the mass itself, as is done
in conventional approaches. This makes the covariant treatment
simple. The COQM has been applied to various problems
\cite{ref4} with satisfactory results. 

Here we would like to note that, when applied to heavy-to-heavy
baryon transitions, our model predictions satisfy the constraints 
imposed by the heavy quark symmetry,\cite{ref6} suggesting that 
COQM is reliable. 
Moreover, our model is also applicable for
heavy-to-light transitions, which are beyond the scope of 
heavy quark effective theory (HQET). 

We organise the paper as follows. Assuming factorization, 
in \S 2 we present the expressions for the decay
rates and the up-down asymmetry parameter $\alpha $.
We use the covariant oscillator quark model to evaluate
the baryonic form factors. Results and discussion are
presented in \S 3.

\section{Methodology}

\subsection{Effective weak Hamiltonian and factorized amplitude}

Neglecting the penguin contribution, the effective
Hamiltonian describing the decays under consideration
is given by

\begin{eqnarray}
&&{\cal H}_{\rm eff} = \frac{G_F}{\sqrt{2}}\; V_{cb}\;V_{q_i q_j}\;
[a_1~ O_1 + a_2~  O_2 ] \label{eq:1}
\end{eqnarray}
with
\begin{eqnarray}
&& O_1=(\bar q_i q_j)^{\mu}\; (\bar c b)_{\mu}~~~~~~~~~~
\mbox{and}~~~~~~~~~~~ O_2=
(\bar q_i b)^{\mu} (\bar c q_j)_{\mu}\;,
\end{eqnarray}
where $a_1$ and $a_2$ correspond to the external and
internal $W$ emission amplitudes. The quark current
$(\bar q_i q_j )_{\mu}=\bar q_i \gamma_\mu(1+\gamma_5) q_j$ denotes
the usual $(V-A)$ current. $q_i$ and $q_j $  are two types
of quark flavors that are hadronized to $P$ or $V$
mesons. Here Eq.~(\ref{eq:1}) is to be understood as an effective
Hamiltonian after considering the Fierz rearrangement.

In the factorization approximation, the baryon decay
amplitude is factorized into a product of two matrix elements
of the form $\langle B^\prime|J_\mu | B \rangle $, which
involves the form factors, and $\langle P(V) |J^\mu |0
\rangle $, which is expressed by the meson decay constant.
The factorized amplitude for the decay processes $\Lambda_b
\to \Lambda_c P(V) $, which proceed via the external
$W$ emission is given as
\begin{equation}
{\cal M}(\Lambda_b \to \Lambda_c P(V) )= \frac{G_F}{\sqrt 2}
V_{cb} V_{q_i q_j}~a_1~\langle P(V)|(\bar q_i q_j)^{\mu}|
0 \rangle \langle \Lambda_c | (\bar c b)_\mu |\Lambda_b
\rangle \;.\label{eq:eqn8}
\end{equation}

Similarly, the amplitude for the decay mode $\Lambda_b
\to \Lambda J/\psi $, which is described by the internal
$W$ emission, is given by Eq. (\ref{eq:eqn8}) in which $a_1$
is replaced by $a_2 $.
The current matrix elements between a pseudoscalar/vector/
axialvector meson\footnote{
We also treat the decay $\Lambda_b\to\Lambda_c+a_1$,
$a_1$ being the axial-vector meson.
} ($P/V/a_1 $) and the vacuum are related
to the corresponding decay constants as
\begin{eqnarray}
&&\langle P(p) |(\bar q_i q_j)^{\mu} |0 \rangle
= -if_P~ p^\mu ,\nonumber\\
&&\langle V(p,\epsilon) |(\bar q_i q_j)^{\mu} |0 \rangle
= M_V~f_V~ \epsilon^\mu ,\nonumber\\
&&\langle a_1(p,\epsilon) |(\bar q_i q_j)^{\mu} |0 \rangle
= M_{a_1}~f_{a_1}~ \epsilon^\mu\;,
\label{eq4}
\end{eqnarray}
where $f_P$, $f_V$ and $f_{a_1}$ are the respective
decay constants. To evaluate the baryon form factors,
we use the covariant oscillator
quark model, which is explicitly presented in the next
section.

\subsection{ Model framework and the Hadronic form factors}

The general treatment of COQM may be called
the ``boosted $LS$-coupling scheme,'' and the wavefunctions, being
tensors in $\tilde U(4) \times O(3,1) $-space, reduce to
those in the $SU(2)_{\rm spin} \times O(3)_{\rm orbit} $-space in the
nonrelativistic quark model in the hadron rest frame. The
spinor and space-time portion of the wave functions separately 
satisfy the respective covariant equations,
the Bargmann-Wigner (BW) equation for the former and the
covariant oscillator equation for the latter. The model parameters 
and form of the wave function are determined completely
through the analysis of mass spectra.\cite{ref3,ref4}

In COQM all the non-exotic $ qqq $ baryons are described\cite{ref6pp} 
by tri-local fields
$ \Phi_{A_1 A_2 A_3} (x_1,x_2,x_3) $,
where the $x_{i}$ are Lorentz four vectors representing the
space time coordinates of constituent quarks,
$A_1=(a,\alpha)\;A_2=(b,\beta)~A_3=(c,\gamma)$, describing the
flavor and covariant spinor of the quarks. The baryon fields
are assumed to satisfy wave equations of the Klein-Gordon type 
and expanded in terms of Fierz components
(that is, the eigenfunctions of ${\cal M}^2$), and are written
as

\begin{equation}
\Phi(X,r \cdots ) =\sum_{P,n}\left (e^{i P_n X}\Psi_n^{(+)}
(r \cdots,P_n)+e^{-iP_n X}\Psi_n^{(-)}(r \cdots ,P_n)\right )\;,
\label{eq:eqn1}
\end{equation}

\begin{equation}
{\cal M}^2(r_{\mu} \cdots , \partial/\partial r_{\mu} \cdots )
\Psi_n^{(\pm)}(r \cdots , P_n)=M_n^2 \Psi_n^{(\pm)}
(r \cdots ,P_n )\;,
\end{equation}
where $X_{\mu}(r_\mu \cdots )$ is the centre-of-mass (relative)
coordinate, and ${\cal M}^2$ is the squared mass operator depending
on relative coordinate variables. The first (second) term of
Eq. (\ref{eq:eqn1}) corresponds to the positive (negative)
frequency part of the centre-of-mass plane-wave motion with
definite total four
momentum $P_n$ and mass $M_n=\sqrt{-P_n^2} $.

The spin portions $U_n$ $(\bar U_n)$ of positive (negative) frequency
internal wave functions $\Psi_n^{(+)}(r \cdots ,P_n) \equiv
U_n(P) f_n(r \cdots, P_n ) $ $(\Psi_n^{(-)} \equiv \bar U_n f_n )$,
satisfy the respective Bargmann-Wigner (BW) equations.
The BW spinor functions are reducible and decompose into the
irreducible components as 
\begin{eqnarray}
U_{n,ABC}&=&\frac{1}{2 \sqrt{2}}[-\gamma_5(1+i v \cdot
\gamma)C^{-1}]_{\alpha \beta}~u_{n,~abc;~\gamma}^{(A)}
+\frac{1}{2 \sqrt{2}}[i\gamma_\mu (1+i v \cdot
\gamma)C^{-1}]_{\alpha \beta} \nonumber\\
& & \times\left [u_{n,~abc;~\gamma,~\mu}
^{(S^*)}+\frac{-1}{\sqrt 3}\left \{(i \gamma_{\mu} +v_{\mu})
\gamma_5~u_{n,~abc}^{(S)}\right \}_\gamma
\right ]\;,\label{eq:eqn2}
\end{eqnarray}
where $v_{\mu}$ is the four velocity of the hadron ($v_{\mu}
=P_{\mu}/M $) and $C$ is the antiparticle conjugate matrix.
The quantities $u_n^{(A)} $, $u_{n, \mu}^{(S^*)} $ and $u_n^{(S)}$ denote
the spin-1/2 `Dirac spinor', spin-3/2 `Rarita-Schwinger
vector-spinor', and the spin-1/2 `Dirac spinor' Fierz components,
respectively. It may be noted that in the $\Lambda_Q$-type
baryons, the two light quarks are in the flavor antisymmetric
and spin-0 state. Thus the BW spinor function for $\Lambda_Q$-type 
baryons is given by the first term of Eq. (\ref{eq:eqn2}).

The oscillator space-time wave function for the ground
state baryons is given\footnote{
In this paper we apply the pure-confining approximation,
neglecting the effect of the one-gluon-exchange potential $U_{\rm OGE}$.
This approximation is expected, aside from the spin-dependent structure,
to be good for the light-light $q\bar q$ and  heavy-light $Q\bar q$ 
meson systems, where the reduced mass of the system is small 
comparatively and the effect of the central potential out of $U_{\rm OGE}$
is expected to be not so large. A similar situation is also expected 
to be valid for the relevant $Qqq$ (or $qqq$) baryon systems.
The well-known phenomenological fact of a linearly-rising Regge 
trajectory for the $q\bar q$ and $qqq$ systems seems to support this 
conjecture. (See the papers referred to in Ref. \citen{ref4}, 
published in 1993 and 1994.)
} by
\begin{equation}
f(P,\rho, \lambda)=f_{\rho}(P;\rho)~f_{\lambda}(P;\lambda),
\end{equation}
where
\begin{equation}
f_{\rho}(P;\rho)=\frac{\beta_{\rho}}{\pi} \exp\left (
-\frac{\beta_\rho }{2}
\left (\rho^2+2\frac{(P\cdot \rho)^2}{M^2}\right ) \right )\;,
\end{equation}
and
\begin{equation}
f_{\lambda}(P;\lambda)=\frac{\beta_{\lambda}}{\pi} \exp\left (
-\frac{\beta_\lambda }{2}
\left (\lambda^2+2\frac{(P\cdot \lambda)^2}
{M^2}\right ) \right )\;,
\end{equation}
with
\begin{equation}
\beta_\rho = \frac{\sqrt{3 m K}}{4}~~~~~~~~~\mbox{and}
~~~~~~~~~~\beta_{\lambda} =\sqrt{\frac{mM K}{2m+M}}\;.\label{eq:eqn3}
\end{equation}
In the above equation, $m$ and $M$ denote the mass of the light
(heavy) quark, and $K$ is the universal spring constant for all
hadronic systems with the value\footnote{
In this paper we apply the values of $\beta$ corresponding to the 
case (A) in Ref. \citen{ref6p}. The values of the branching ratios
with $\beta$ corresponding to the case (B) generally become
slightly smaller  (by $\stackrel{<}{\sim}20\%$).
} $K$ =0.106 $\mbox{GeV}^3 $.\cite{ref8} The internal
relative coordinates are defined as
\begin{equation}
\rho_{\mu}=x_{2 \mu}-x_{1 \mu}~~~~~~~\mbox{and}\;\;\;\;\;\;
\lambda_{\mu}=x_{3 \mu} - \frac{x_{1\mu}+x_{2 \mu}}{2}\;.
\end{equation}

The effective action for weak interactions of baryons
with $W$-bosons is given by

\begin{eqnarray}
S_W &=& \int d^4 x_1 d^4 x_2 d^4 x_3~ \langle \bar \Phi_{F, P^\prime}
(x_1,x_2,x_3)^{C^\prime B A}~ i\gamma_{\mu}
(1+\gamma_5{)_{\gamma^\prime}}^{\gamma}\nonumber\\
& & \times  \Phi_{I, P}(x_1,x_2,x_3)_{ABC}
\rangle W_{\mu,q}(x_3)\;,
\end{eqnarray}
where we have omitted the CKM matrix elements and the coupling
constants. This is obtained from the consideration of covariance,
assuming a quark current with the standard $V-A $ form.
Here $\Phi_{I,P}~(\bar \Phi_{F,P^\prime}) $ denotes
the initial (final) baryons with definite four momentum
$P_\mu~(P_{\mu}^{\prime}) $, and $q_\mu $ is the momentum of
the $W$ boson. The notation $\langle~~\rangle $ represents 
the trace of Dirac
spinor indices. Concentrating only on $\Lambda_Q (Qud)
\to \Lambda_{Q^\prime} (Q^\prime ud) $, the relevant effective
currents $J_{\mu}(X)_
{P^\prime,P} $ are obtained by identifying the above action
with
\begin{equation}
S_W =\int d^4 X J_{\mu}(X)_{P^\prime,P}\;W_{\mu}(X)_q\;.
\end{equation}
Then $J_{\mu}(X=0)_{P^\prime,P} \equiv J_\mu $
is explicitly given as,\cite{ref6pp}

\begin{equation}
J_{\mu}^{Q Q^\prime}=I_{ud}^{Q^\prime Q}(w)~
\bar u_A^{Q^\prime q_2 q_1}(v^\prime)~\frac{w+1}{2}~
i \gamma_{\mu}(1+\gamma_5)~u_{A,Qq_1 q_2 }(v)\;.
\label{eq15}
\end{equation}
The quantity $I_{ud}^{Q^\prime Q}(w) (w\equiv -v\cdot v')$, 
the overlapping of the initial and final wave functions, 
represents the universal form factor. 
It describes the confined effects of quarks and is given by

\begin{equation}
I_{ud}^{Q^\prime Q}(w)=\frac{1}{w}~\frac{4 \beta_{\lambda}
\beta_{\lambda}^\prime}
{\beta_{\lambda}+\beta_{\lambda}^\prime }
\frac{1}{\sqrt{C(w)}} \exp(-G(w))\;,
\label{eqnI}
\end{equation}
where
\begin{equation}
C(w)=(\beta_\lambda-\beta_\lambda^{\prime})^2+
4\beta_\lambda \beta_\lambda^{\prime}~w^2\;,
\end{equation}
and
\begin{equation}
G(w)= \frac{4 m_q^2 (\beta_\lambda+\beta_\lambda^\prime)~w (w-1)}
{(\beta_\lambda - \beta_\lambda^\prime )^2 + 4
\beta_\lambda \beta_\lambda^\prime~w^2}\;.
\end{equation}
The expression for $\beta_\lambda $ is given by Eq.
(\ref{eq:eqn3})

The form factor function $I_{ud}^{cb}(w)$ for
$ \Lambda_b \to \Lambda_c $ decays
corresponds to the baryonic Isgur-Wise function $\eta(w) $ in 
HQET.\cite{ref6} At the zero recoil point $w=1$,
the value of $I_{ud}^{cb}(w)$ is given
by
\begin{equation}
I_{ud}^{cb}(w=1)=\frac{1}{w}\frac{4 \beta_{\lambda}
\beta_{\lambda}^\prime}
{(\beta_{\lambda}+\beta_{\lambda}^\prime)^2}\;.\label{eq:eqn7}
\end{equation}
In the heavy quark symmetry limit we have $\beta_\lambda=
\beta_{\lambda}^\prime $, so that 
Eq. (\ref{eq:eqn7}) correctly reproduces\cite{ref6p} 
the normalization condition of
HQET, i.e., $\eta(w=1)=1 $. However, HQET as it is predicts
nothing about the Isgur-Wise function except for the zero recoil
point, while in COQM the form factor functions can be derived
at any kinematical point of interest.
In addition, the functional form Eq.~(\ref{eqnI}) of
the COQM form factor $I_{ud}^{Q^\prime Q}(w)$
also applies to the heavy-to-light transition processes,
whereas HQET does not provide anything for this sector.

\subsection{Decay rates and aymmetry parameters}

After obtaining the effective current in the COQM
with Eqs.~(\ref{eq:eqn8}) and (\ref{eq15}), 
one can write the
transition amplitude for the decay mode $\Lambda_b \to
\Lambda_c P $ as
\begin{eqnarray}
{\cal M}(\Lambda_b(v) \to \Lambda_c(v^\prime) P(p))&=&
\frac{G_F}{2\sqrt 2} V_{cb} V_{q_i q_j}~ a_1~f_P~p^{\mu}~I_{ud}^{cb}
(w)~(w+1)\nonumber\\
& & \times \bar u_A^{cud}(v^\prime)~
\gamma_{\mu}(1+\gamma_5)~u_{A,bud}(v)\;,\label{eq:eqn4}
\end{eqnarray}
and the corresponding decay width as
\begin{eqnarray}
\Gamma (\Lambda_b \to \Lambda_c P )&=& \frac{G_F^2}{32 \pi
M_{\Lambda_b}^2} |V_{cb} V_{q_i q_j}|^2~a_1^2~ f_P^2~
(I_{ud}^{cb}(w))^2 (w+1)^2~|{\bf p}|\nonumber\\
& & \times  \left [(M_{\Lambda_b}^2 -M_{\Lambda_c}^2)^2
-M_P^2(M_{\Lambda_b}^2+M_{\Lambda_c}^2)\right ]\;,
\end{eqnarray}
where ${\bf p}$ is the c.m.~momentum of the emitted particles
in the rest frame of the parent $\Lambda_b $ baryon.

To obtain the asymmetry parameter $\alpha $, we write
Eq. (\ref{eq:eqn4}), substituting
$p^\mu=M_{\Lambda_b} v^{\mu} - M_{\Lambda_c} v^{\prime \mu} $
as
\begin{equation}
{\cal M}(\Lambda_b \to \Lambda_c P)=f_P~ \bar u_{A}^{cud}
~(G_1+\gamma_5G_2)~u_{A,bud}(v)\;,
\end{equation}
where
\begin{equation}
G_1=\lambda(M_{\Lambda_b}-M_{\Lambda_c})\;,~~~~~~~~~~~
G_2=-\lambda(M_{\Lambda_b}+M_{\Lambda_c})\;,
\end{equation}
with
\begin{equation}
\lambda=\frac{G_F}{2 \sqrt 2} V_{cb} V_{q_i q_j} ~a_1
~I_{ud}^{cb}(w)~ (1+w)\;.
\end{equation}
In terms of the new form factors $G_1$ and $G_2$,
the asymmetry parameter $\alpha$ is given by \cite{ref91}
\begin{equation}
\alpha=\frac{2 G_1 G_2 |{\bf p}|}{(E_{\Lambda_c}+M_{\Lambda_c})
G_1^2+(E_{\Lambda_c}-M_{\Lambda_c})G_2^2}\;.
\end{equation}

To obtain the decay rate for the $\Lambda_b \to \Lambda_c V $
transition, we write the transition amplitude for
the process as
\begin{eqnarray}
{\cal M}(\Lambda_b(v) \to \Lambda_c(v^\prime)
V(p,\epsilon))&=&
i\frac{G_F}{2\sqrt 2} V_{cb} V_{q_i q_j}~ a_1~f_V~M_V~
\epsilon^{\mu}~I_{ud}^{cb}
(w)~(w+1)\nonumber\\
& & \times \bar u_A^{cud}(v^\prime)~
\gamma_{\mu}(1+\gamma_5)~u_{A,bud}(v)\;.\label{eq:eqn5}
\end{eqnarray}
The corresponding decay width is given by
\begin{eqnarray}
\Gamma (\Lambda_b \to \Lambda_c V )&=& \frac{G_F^2}{32 \pi
M_{\Lambda_b}^2} |V_{cb} V_{q_i q_j}|^2~a_1^2~ f_V^2~
(I_{ud}^{cb}(w))^2 (w+1)^2~|{\bf p}|\nonumber\\
& & \times  \left [(M_{\Lambda_b}^2 -M_{\Lambda_c}^2)^2
+M_V^2(M_{\Lambda_b}^2+M_{\Lambda_c}^2-2M_V^2)\right ]\;.
\label{eqn27}
\end{eqnarray}
To obtain the asymmetry parameter, we compare Eq.
(\ref{eq:eqn5}) to the general form \cite{ref09}
in the rest frame of $\Lambda_b $,
\begin{equation}
\chi_{\Lambda_c}^{\dagger}[ S {\bf \sigma} +P_1 {\hat{\bf p}}
+i P_2 ({\hat{\bf p}} \times {\bf \sigma})+D({\bf \sigma}\cdot
{\hat{\bf p}}) {\hat{\bf p}}]
\cdot {\bf \epsilon}~ \chi_{\Lambda_b}\;,
\end{equation}
with ${\hat{\bf p}}$ now a unit vector in the direction
of the $\Lambda_c $ baryon.
The values for the four amplitudes $ S,~ P_1,~P_2$ and $D$
are given as
\begin{equation}
S=i \frac{G_F}{2 \sqrt 2}~ V_{cb} V_{q_i q_j}~ a_1~ 
f_V~M_V~(I_{ud}^{cb}(w))(1+w)\;,
\end{equation}
\begin{equation}
P_1/S= - \left ( \frac{M_{\Lambda_b}+M_{\Lambda_c}}
{E_{\Lambda_c}+M_{\Lambda_c}} \right )
\left ( \frac{|{\bf p}|}{E_V} \right ) ,
\end{equation}
\begin{equation}
P_2/S=  \left ( \frac{|{\bf p}| }
{E_{\Lambda_c}+M_{\Lambda_c}} \right ) ,
\end{equation}
\begin{equation}
D/S=  \left ( \frac{|{\bf p}|^2 }
{E_V(E_{\Lambda_c}+M_{\Lambda_c})} \right ) .
\end{equation}

From these expressions, the up-down asymmetry $\alpha $
of the final $\Lambda_c $ with respect to $\Lambda_b $
polarization is given by
\begin{equation}
\alpha=2\mbox{Re}~\frac{[(1+D/S)^*P_1/S+2(P_2^*/S) M_V^2/E_V^2]}
{K}\;,
\end{equation}
where
\begin{equation}
K=[|1+D/S|^2+|P_1/S|^2+2(1+|P_2/S|^2)M_V^2/E_V^2]\;.
\end{equation}
The decay rate for $\Lambda_b \to \Lambda J/\psi $ is given
in analogy to Eq. (\ref{eqn27}) with $a_1$ replaced by $a_2$ and
$f_V $ replaced by $f_{J/\psi}$, with the relevant CKM matrix elements
as $|V_{cb} V_{cs}|^2$.

The COQM is applicable not only for heavy-to-heavy transitions 
but also for heavy-to-light transitions, where the final baryon
moves extremely relativistically, as was mentioned in \S 1. Here
we consider the case of the decay mode $\Lambda_b
\to \Lambda \bar D^0 $ to obtain the CKM-Wolfenstein
parameter $(\rho^2+\eta^2)$ from the ratio of the
the decay widths $\Gamma (\Lambda_b \to \Lambda \bar D^0) /
\Gamma (\Lambda_b \to \Lambda J/\psi)$. The decay width
for the process $\Lambda_b \to \Lambda \bar D^0 $ is given as
\begin{eqnarray}
\Gamma (\Lambda_b \to \Lambda \bar D^0 )&=& \frac{G_F^2}{32 \pi
M_{\Lambda_b}^2} |V_{ub} V_{cs}|^2~a_2^2~ f_D^2~
(I_{ud}^{sb}(w))^2 (w+1)^2~|{\bf p}|\nonumber\\
& & \times \left [(M_{\Lambda_b}^2 -M_{\Lambda}^2)^2
-M_D^2(M_{\Lambda_b}^2+M_{\Lambda}^2)\right ]\;.
\end{eqnarray}

\section{Results and conclusion}

\begin{table}
\caption{Branching ratios (in percent) for different
nonleptonic $\Lambda_b$ decay
processes and their asymmetry parameters $\alpha$.}
\vspace {0.2 true in}
\begin{center}
\begin{tabular}{lllll}
\hline
\multicolumn{1}{c}{Decay processes}&
\multicolumn{1}{c}{$\alpha $} &
\multicolumn{1}{c}{Br. ratio in $\%$}\\
\hline
$\Lambda_b \rightarrow \Lambda_c^+ \pi^-$ &
$-0.999$ & 0.175\\
$\Lambda_b \rightarrow \Lambda_c^+ K^-$ &
$-1.000$ & 0.013\\
$\Lambda_b \rightarrow \Lambda_c^+ D^-$ &
$-0.987$ & 0.030 \\
$\Lambda_b \rightarrow \Lambda_c^+ D_s^-$ &
$-0.984$& 0.77 \\
$\Lambda_b \rightarrow \Lambda_c^+ \rho^-$ &
$-0.898 $  & 0.491 \\
$\Lambda_b \rightarrow \Lambda_c^+ K^{*-}$ &
$-0.865$ & 0.027 \\
$\Lambda_b \rightarrow \Lambda_c^+ D^{*-}$ &
$-0.459$ & 0.049\\
$\Lambda_b \rightarrow \Lambda_c^+ D_s^{*-} $ &
$-0.419$ & 1.414  \\
$\Lambda_b \to \Lambda_c^+ a_1 $& $-0.758 $&
0.532 \\
$\Lambda_b \rightarrow \Lambda J/\psi$&
$-0.208$ & $2.55 \times 10^{-2}$\\
% modifed at isida lab. " " -> "%" for tex compile.
\hline
\end{tabular}
\end{center}
\end{table}

\begin{table}
\caption{Branching ratios (in percent) for different $\Lambda_b$
processes and comparison with other calculations.}
\vspace {0.2 true in}
\begin{center}
\begin{tabular}{llllll}
\hline
\multicolumn{1}{c}{Decay processes}&
\multicolumn{1}{c}{This work}&
\multicolumn{1}{c}{Ref. \citen{ref05} }&
\multicolumn{1}{c}{Ref. \citen{ref06}} &
\multicolumn{1}{c}{Ref. \citen{ref07} } &
\multicolumn{1}{c}{Ref. \citen{ref08} }\\
\multicolumn{1}{c}{}&
\multicolumn{1}{c}{}&
\multicolumn{1}{c}{}&
\multicolumn{1}{c}{Large $N_c$} &
\multicolumn{1}{c}{Large $N_c$} &
\multicolumn{1}{c}{Large $N_c$}\\
\hline
$\Lambda_b \rightarrow \Lambda_c^+ \pi^-$ &
0.175 &0.38  & - &0.391 & 0.503\\
$\Lambda_b \rightarrow \Lambda_c^+ K^-$ &
0.013 & - &- &- & 0.037 \\
$\Lambda_b \rightarrow \Lambda_c^+ D^-$ &
0.030 &-&- &- &- \\
$\Lambda_b \rightarrow \Lambda_c^+ D_s^-$ &
0.77&1.1 &2.23& 1.291 &- \\
$\Lambda_b \rightarrow \Lambda_c^+ \rho^-$ &
0.491&0.54 & -  &1.082  &0.723 \\
$\Lambda_b \rightarrow \Lambda_c^+ K^{*-}$ &
0.027 & - &- &- & 0.037 \\
$\Lambda_b \rightarrow \Lambda_c^+ D^{*-}$ &
0.049&- &-&- &- \\
$\Lambda_b \rightarrow \Lambda_c^+ D_s^{*-} $ &
1.414&0.91&3.26 & 1.983 & - \\
$\Lambda_b \to \Lambda_c a_1 $ & 0.532 &-&-&-&-\\
$\Lambda_b \rightarrow \Lambda J/\psi$&
$2.49 \times 10^{-2}$ &$1.6 \times 10^{-2} $
& $6.037 \times 10^{-2}$ &-&-\\
\hline
\end{tabular}
\end{center}
\end{table}

\begin{table}
\caption{Asymmetry parameter $\alpha$ for different $\Lambda_b$
processes and comparison with other calculations.}
\vspace {0.2 true in}
\begin{center}
\begin{tabular}{lllll}
\hline
\multicolumn{1}{c}{Decay processes}&
\multicolumn{1}{c}{This work}&
\multicolumn{1}{c}{Ref. \citen{ref05} }&
\multicolumn{1}{c}{Ref. \citen{ref06}} &
\multicolumn{1}{c}{Ref. \citen{ref08}}\\
\multicolumn{1}{c}{}&
\multicolumn{1}{c}{}&
\multicolumn{1}{c}{}&
\multicolumn{1}{c}{Large $N_c$}&
\multicolumn{1}{c}{Large $N_c$} \\
\hline
$\Lambda_b \rightarrow \Lambda_c^+ \pi^-$ &
$-0.999$ &$-0.99$ & - &$-1.00$\\
$\Lambda_b \rightarrow \Lambda_c^+ K^-$ &
$-1.000$ & - & -&$-1.00$ \\
$\Lambda_b \rightarrow \Lambda_c^+ D^-$ &
$-0.987$&-&- &-  \\
$\Lambda_b \rightarrow \Lambda_c^+ D_s^-$ &
$-0.984$& $-0.99 $&$-0.98$&- \\
$\Lambda_b \rightarrow \Lambda_c^+ \rho^-$ &
$-0.898$ &$-0.88$ & - &$- 0.885$ \\
$\Lambda_b \rightarrow \Lambda_c^+ K^{*-}$ &
$-0.865$ & - & - &$-0.885$  \\
$\Lambda_b \rightarrow \Lambda_c^+ D^{*-}$ &
$-0.459$&- &- &- \\
$\Lambda_b \rightarrow \Lambda_c^+ D_s^{*-} $ &
$-0.419$ &$-0.36 $&$-0.40$ & -  \\
$\Lambda_b \to \Lambda_c a_1 $ & $-0.758 $&
-& -& -\\
$\Lambda_b \rightarrow \Lambda J/\psi$&
$-0.208$ &$-0.1$& $-0.18$ &-\\
\hline
\end{tabular}
\end{center}
\end{table}

In order to make a numerical estimate, we use the
following values of various quantities. The quark masses
(in GeV) are taken as $m_u=m_d=0.4$, $m_s=0.51$, $m_c=1.7$
and $m_b=5$, which are determined from the analysis\cite{refmR}
of meson mass spectra. 
The particle masses and lifetimes are taken from Ref.
\citen{ref10}. The relevant CKM parameters\cite{ref10} 
used are $V_{cb}=0.0395$,
$V_{cs}=1.04 $, $V_{cd}=0.224$, $V_{ud}=0.974$ and $V_{us}
=0.2196$. The decay constants are taken as $f_{\pi}=130.7$,
$f_K=159.8$; $f_{K^*}=214$;\cite{refa} $f_{\rho}=210$;\cite{refb}
$f_D=220$, $f_{D^*}=230$, $f_{D_s}=240$, $f_{D_s^*}=260$\cite{refc} 
and $f_{a_1}=205$\cite{refd} (in MeV).
The decay constant $f_{J/\psi}$ is determined from the value of
$\Gamma(J/\psi \to e^+ e^- )$:\cite{ref10}
\begin{equation}
f_{J/\psi}=\sqrt{\frac{9}{4} \left (\frac{3}{4\pi \alpha^2}
\right ) \Gamma(J/\psi \to e^+ e^-)M_{J/\psi}}=404.5 ~\mbox{MeV}\;.
\end{equation}

The parameters $a_1$ and $a_2$ appearing in these decays have
recently been extracted from the CLEO data and turn out to be
$a_1=1.05$ and $a_2=0.25$.\cite{ref02}
Using these values we have obtained the values of the branching
ratios and the asymmetry parameter $\alpha$ for
several nonleptonic $\Lambda_b $ decays. These values are tabulated
in Table I. The asymmetry parameter $\alpha $ in all
these decay modes is found to be negative, which indicates
the $V-A$ nature of the weak current.
Our predicted branching ratio for the decay process
$\Lambda_b \to \Lambda J/\psi $ ($2.55 \times 10^{-4})$
is consistent with the recent experimental data \cite{ref10}
($4.7 \pm 2.7 \times 10^{-4}$): Here it may be worthwhile to note
that the final $\Lambda$ in this process experiences relativistic motion
with velocity $v_\Lambda =0.84c$, and the form factor function
plays a significant role taking a value $I(w)=0.11$.

From the ratio of the decay widths $\Lambda_b \to
\Lambda \bar D^0 $ and $\Lambda_b \to \Lambda J/\psi $, we obtain
\begin{equation}
\frac{\Gamma(\Lambda_b \to \Lambda \bar D^0)}{\Gamma
(\Lambda_b \to \Lambda J/\psi)}=10.235 \times 10^{-2}~
|V_{ub}/V_{cb}|^2=4.936 \times 10^{-3}~
(\rho^2 +\eta^2 )\;.\label{eq:eqn9}
\end{equation}
Substituting the value $|V_{ub}/V_{cb} |=0.08\pm 0.02$, which
is measured in charmless $b$ decays,\cite{ref03}
into this relation, we obtain from Eq. (\ref{eq:eqn9})
\begin{equation}
(\rho^2+\eta^2)^{1/2}=0.364\pm 0.091,
\end{equation}
which is in excellent agreement with the recent prediction
\cite{ref04} $(\rho^2+\eta^2)^{1/2} =0.36 \pm 0.09 $.

In this paper we have calculated the branching ratios of
the exclusive nonleptonic decays of $\Lambda_b$ baryons using the
covariant oscillator quark model, on the basis of the factorization
approximation. The manifestly covariant weak currents are given
by the overlapping integrals between the initial and final
hadron wave functions. These currents are represented by a
common form factor function, and for heavy $\to $ heavy
baryon transitions this form factor function corresponds
to the Isgur-Wise function of HQET and has similar properties
at the zero recoil point. Using these currents we have derived
the decay rates and up-down asymmetry parameter $\alpha $
for various nonleptonic weak decays of $\Lambda_b $
baryons. Our predicted result for the branching ratio
$Br (\Lambda_b \to \Lambda J/\psi )$ is in agreement with the
currently available experimental data.
Recently, these decay processes have been studied using the
quark model \cite{ref05,ref06} and HQET\cite{ref07,ref08} 
in the large $N_c $ limit.
However, our predicted results for the branching ratios
are smaller than the previous values, as can be seen from Table II.
Future experimental data from the colliders are expected
to verify and distinguish the various results.
However, the values of the asymmetry parameter 
in all these calculations are
nearly the same. Furthermore, the CKM-Wolfenstein parameter
$(\rho^2+\eta^2)^{1/2}$ obtained in the framework of our
model agrees very well with the recent prediction.

\acknowledgements

R. M. would like to thank CSIR, Govt. of India, for a fellowship.
A. K. G. and M. P. K. acknowledge financial support from
DST, Govt. of India.

\end{document}